\documentclass[a4paper,10pt]{article}
\usepackage[latin1]{inputenc}
\usepackage[dvips]{graphicx}
\usepackage{amssymb}

\newcommand{\ve}{\varepsilon}

\begin{document}

\twocolumn[\hsize\textwidth\columnwidth\hsize\csname@twocolumnfalse%
\endcsname

\noindent
{\Large\bf Bias dependent subband edges and the 0.7 
conductance anomaly} 

\vspace*{2mm} \noindent
{\large Anders Kristensen$^\dagger$ and Henrik Bruus$^\ddagger$}

\vspace*{2mm} \noindent
{$^\dagger$Mikroelektronik Centret, Build.\ 345, Technical
  University of Denmark, DK-2800 Lyngby, Denmark} 

\noindent
{$^\ddagger$Niels Bohr Institute, {\O}rsted Laboratory,
  Universitetsparken 5, DK-2100 Copenhagen, Denmark}

\vspace*{1mm} \noindent
{\em Accepted for publication in
  Physica Scripta (2002), (Proc.\ 19th Nordic Semiconductor Meeting, 2001)}

\vspace*{1mm} \noindent
{PACS Ref: 73.61.-r, 73.23.-b, 73.23.Ad}
\vspace*{15mm}

]

\noindent
{\bf Abstract}\\[3mm]
\begin{minipage}{\columnwidth}
\setlength{\baselineskip}{0.8\baselineskip}
{\small  
The 0.7~$(2e^2/h)$ conductance anomaly is studied in strongly
confined, etched GaAs/GaAlAs quantum point contacts by measuring the
differential conductance $G$ as a function of source-drain bias 
$V_{\rm sd}$ and gate-source bias $V_{\rm gs}$ as well as a
function of temperature. In the $V_{\rm gs}$--$V_{\rm sd}$ plane
we use a grayscale plot of the transconductance $dG/dV_{\rm gs}$ to
map out the bias dependent transitions between the normal and
anomalous conductance plateaus. Any given transition is interpreted
as arising when the bias controlled chemical potential $\mu_d$
($\mu_s$) of the drain (source) reservoir crosses a subband edge
$\ve_\alpha$ in the point contact. From the grayscale plot we extract the
constant normal subband edges $\ve_0$, $\ve_1$, $\ldots$  and most
notably the bias dependent anomalous subband edge $\ve'_0(\mu_d)$
split off from  $\ve_0$. We show by applying a finite--bias version of
the recently proposed BCF model, how the bias dependence of the
anomalous subband edge is the key to analyze various experimental
observations related to the 0.7 anomaly.  
}
\end{minipage}
\vspace*{3mm}

\noindent
{\bf 1. Introduction}\\[2mm]
The quantized conductance of a narrow quantum point contact with
subband edges $\ve_\alpha$ connecting the source and drain reservoirs of
two-dimensional electron gas (2DEG) was discovered in 1988 by van Wees
et al.~\cite{vanWees88} and Wharam et al.~\cite{Wharam88}. The
quantization of the conductance in units of the spin degenerate
conductance quantum, $G_2 = 2\:e^2/h$, can be explained within a
single-particle picture in terms of the Landauer-B\"{u}ttiker
formalism involving the transmission coefficients 
${\cal T}[\ve,\ve_\alpha]$ of the subbands $\alpha$ in the
constriction. For a review see Ref.~\cite{vanHouten92}. 

\begin{figure}[tbp]
\centerline{
  \includegraphics[height=80mm]{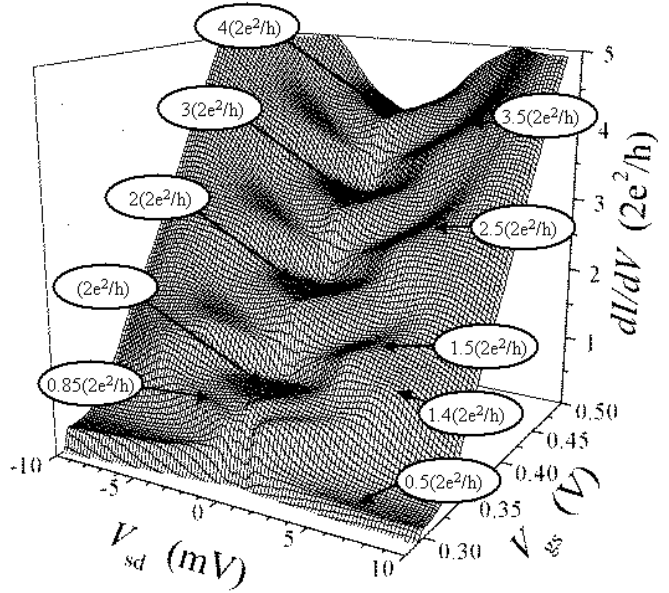}}
\caption{\label{fig:gatebias_surface}
Measurement of the differential conductance 
$G = dI/dV_{\rm sd}$ versus gate-source
voltage $V_{\rm gs}$ and source-drain bias voltage $V_{\rm sd}$.
The data are taken at $T = 0.3$~K on sample A in
Ref.~\cite{Kristensen00}. Normal integer conductance plateaus, $G =
n\: 2e^2/h$, are seen at $V_{\rm sd} = 0$~mV and the well-known
half-plateaus, $G = (n+\frac{1}{2})\: 2e^2/h $, at $V_{\rm   sd} \sim
\pm3$~mV. In addition anomalous plateaus appear with 
$ G = 0.85\: 2e^2/h $ and $G = 1.4\: 2e^2/h$ at $V_{\rm sd} \sim \pm2$~mV
and $\sim \pm7$~mV, respectively. 
}
\end{figure}

The observation by Thomas et al.~\cite{Thomas96}, later confirmed
by others \cite{Tscheuschner96}--\cite{Reilly00}
of the so-called 0.7 conductance anomaly is possibly the first
experimental indication of many-body phenomena in semiconductor
quantum point contacts and wires. At zero source-drain bias, 
$V_{\rm sd}$, the 0.7 anomaly is a suppression of the differential
conductance, $G = dI/dV_{\rm sd}$, at the low-density end of the first
quantized conductance plateau, $G = 1.0\: G_2$. It appears as a
shoulder-like structure around $G = 0.7\: G_2$, which sometimes
develops into a conductance plateau, that becomes more pronounced as
the temperature is raised. At high temperatures the conductance
suppression exhibits activated behavior as function of temperature
\cite{Kristensen98b,Kristensen00}, and a density-dependent activation
temperature, $T_A$, or energy, $\Delta = k T_A$ can be
extracted. At finite $V_{\rm sd}$, the 0.7 anomaly evolves into a
well-defined conductance plateau at $G = 0.85\: G_2$
\cite{Thomas98b,Thomas98} and, as $V_{\rm sd}$ is increased further,
into another plateau at $G = 1.4\: G_2$ \cite{Kristensen00}, see also
Fig.~\ref{fig:gatebias_surface}.  

In addition to the differential conductance measurements, the 0.7
anomaly has also been studied using other experimental methods: the
series conductance of two point contacts \cite{Liang99}, suppression
of shot-noise \cite{Griffiths00}, thermopower measurements
\cite{Appleyard} and acousto-electric currents induced by surface
acoustic waves (SAW) \cite{Cunningham2001}. 

Recently Bruus, Cheianov, and Flensberg proposed a model
\cite{Bruus01}, henceforth denoted the BCF model, which provides 
a unifying explanation of different types of experiments related to
the 0.7 anomaly: the anomalous effects are interpreted in terms of
anomalous, bias dependent 1D subband edges, split off from
each normal 1D subband. In our opinion, further investigations of
these anomalous subband edges is the key to a deeper
understanding of the 0.7 structure. This point of view will be
emphasized in this paper by comparing the results of a number of
experiments with model calculations based solely on the input 
from one experiment: the low-temperature determination of the
anomalous subband edge $\ve'(\mu_d)$.
\vspace*{3mm}

\noindent
{\bf 2. The bias dependent subband edge 
\boldmath{$\ve'(\mu_d)$}}\\[2mm]
The experimental data presented in this paper are all obtained from
sample A described in detail in our previous paper
\cite{Kristensen00}, a shallow etched topgated GaAs/GaAlAs quantum
point contact 200~nm wide opening (determined lithographically).
An overview of the observed conductance quantization is presented in
the surface plot Fig.~\ref{fig:gatebias_surface}, where the
differential conductance $G = dI/dV_{\rm sd}$ is plotted versus  
gate-source voltage $V_{\rm gs}$ and source-drain bias voltage 
$V_{\rm sd}$. The data are taken at $T = 0.3$~K. Normal integer
conductance plateaus \cite{vanWees88,Wharam88}, $G = n\: G_2$, are
seen at $V_{\rm sd} = 0$~mV and the well-known half-plateaus 
\cite{Patel91,Glazman89}, $G = (n+\frac{1}{2})\: G_2 $, at  
$V_{\rm sd} \sim \pm3$~mV. In addition anomalous plateaus 
\cite{Kristensen00,Thomas98b,Thomas98} appear with 
$G = 0.85\: G_2 $ and $G = 1.4\: G_2$ at $V_{\rm sd} \sim \pm2$~mV and
$\sim \pm7$~mV, respectively. A thorough analysis of this data,
including e.g.\ electrostatic self-gating effects, is given in Sec.~IV
of Ref.~\cite{Kristensen00}. In this paper we focus on the anomalous
$G = 0.85\: G_2$ and $1.4\: G_2$ plateaus.

\begin{figure}[tbp]
\centerline{
  \includegraphics[height=80mm]{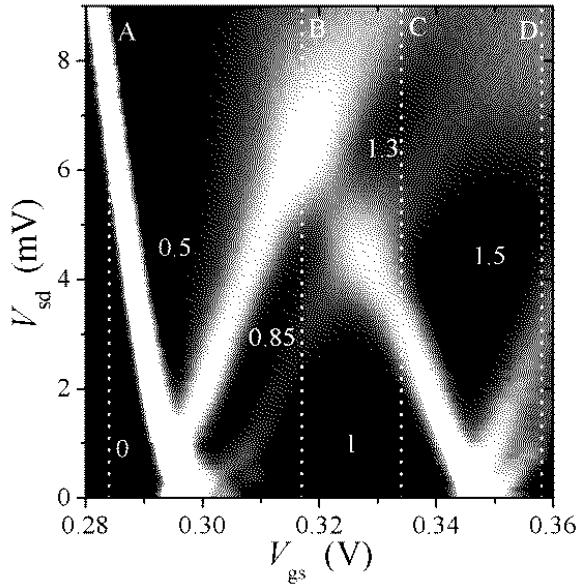}}
\caption{\label{fig:grayscaleEXP}
Part of the experimental data from Fig.~\ref{fig:gatebias_surface}
presented as a grayscale plot of the transconductance, 
$dG/dV_{\rm gs}(V_{\rm gs}, V_{\rm sd})$. Black corresponds to zero
transconductance, i.e.\ to plateaus in the differential conductance, 
$G=dI/dV_{\rm sd}$. White corresponds to high transconductance. The
bright lines in the plot therefore indicate the positions $(V_{\rm gs},
V_{\rm sd})$ of transitions between the various conductance
plateaus. The numbers indicate the value of $G$ in units of $2e^{2}/h$
on the various plateaus.  The dashed vertical lines A--D indicate
sweeps in $V_{\rm sd}$ at four fixed values of $V_{\rm gs}$; see
Sec.~5. 
}
\end{figure}

The conductance anomalies appear very clearly in the transconductance,
$dG/dV_{\rm gs}$, which is calculated by numerical differentiation of
the data in Fig.~\ref{fig:gatebias_surface}. In
Fig.~\ref{fig:grayscaleEXP} it is represented as a grayscale plot in
the  $V_{\rm gs}$--$V_{\rm sd}$ plane. Plateau regions, e.g.\
regions with small transconductance appear as black regions bounded by
bright, high transconductance transition regions. The clearest
feature of the plot is the usual straight but slanted white transition
lines surrounding the integer and half-integer plateaus $G = 0.0, 0.5,
1.0$, and $1.5\:G_2$ \cite{Thomas98}. But the feature of interest in
this paper is the weaker gray transition line dividing the $G=0.85\: G_2$
plateau from the $G = 1.0\: G_2$ plateau and further on dividing the
$G=1.4\: G_2$ plateau from the $G= 1.5\: G_2$ plateau. 

The central point in our analysis is that we interpret all
transitions between conductance plateaus, including the anomalous
ones, as arising when the chemical potential $\mu_s$ of the source or
$\mu_d$ of the drain reservoir crosses the subband edges $\ve_\alpha$
in the point contact.  
       
In the conventional half-plateau analysis 
\cite{Patel91,Glazman89} the straight white transition lines between
the conductance plateaus $G =$ 0.0, 0.5, 1.0, and $1.5\: G_2$ are
explained as follows. For $V_{\rm sd} = 0$~mV the $G = 1.0\: G_2$
plateau begins at $V_{\rm gs} = V_{\rm gs}^a$ and ends at  
$V_{\rm gs} = V_{\rm gs}^b$,
\begin{equation} \label{VaVb}
V_{\rm gs}^a = 292~{\rm mV}, \qquad
V_{\rm gs}^b = 347~{\rm mV}.
\end{equation}
Each value of $V_{\rm gs}$ results in a certain average chemical
potential $\mu_0(V_{\rm gs})$. We shall assume a simple linear
function with coefficients $\alpha$ and $\beta$ to be determined
shortly, 
\begin{equation} \label{mu0_V}
\mu_0(V_{\rm gs}) = \alpha V_{\rm gs} + \beta.
\end{equation}
In energy space the $G = 1.0\: G_2$ plateau occurs when the chemical
potential lies between the subband edges $\ve_0$ and $\ve_1$, and we
have  
\begin{equation} \label{mu0amu0b}
\mu_0(V_{\rm gs}^a) \equiv \ve_0, \qquad
\mu_0(V_{\rm gs}^b) \equiv \ve_1.
\end{equation}
As finite source-drain bias is applied the chemical potentials $\mu_s$
and $\mu_d$ of the source and drain reservoirs are shifted away from
their average value $\mu_0$ according to
\begin{eqnarray} \label{mu_s_mu_d}
\mu_d(V_{\rm gs},V_{\rm sd}) &\equiv& 
\mu_0(V_{\rm gs}) - \frac{1}{2}\:eV_{\rm sd},\\
\mu_s(V_{\rm gs},V_{\rm sd}) &\equiv& 
\mu_0(V_{\rm gs}) + \frac{1}{2}\:eV_{\rm sd}.
\end{eqnarray}
In the following we take $V_{\rm sd}>0$ and denote the reservoir with the
lower chemical potential the drain reservoir. A transition from one
plateau to another happens when $\mu_d$ aligns with a subband edge
$\ve_n$ or when $\mu_s$ aligns with $\ve_m$, 
\begin{eqnarray} 
\label{transition_d}
\mu_d =  \ve_n &\! \Rightarrow\! & 
V_{\rm sd} = \frac{2}{e}(\beta-\ve_n+\alpha V_{\rm gs}),\\
\label{transition_s}
\mu_s =  \ve_m &\! \Rightarrow\! & 
V_{\rm sd} = \frac{2}{e}(\ve_m-\beta-\alpha V_{\rm gs}).
\end{eqnarray}
Clearly, transition lines with positive (negative) slope in the
grayscale plot occur when the chemical potential in the drain (source)
crosses a subband edge $\ve_n$ ($\ve_m$). We can immediately conclude
that the anomalous transition line with its positive slope involves
the splitting off of the anomalous subband edge $\ve'_0$ from
the normal subband edge $\ve_0$ when the drain chemical potential
$\mu_d$ crosses $\ve_0$, while the subband remains unchanged as
the source chemical potential $\mu_s$ crosses $\ve_0$.

There is one fundamental difference between the normal and the
anomalous plateaus. At finite bias normal spin-degenerate plateaus
occurs in intervals of $1/2\: G_2$, while spin polarized plateaus
occurs in intervals of $1/4\: G_2$. The anomalous ones involves units 
of $1/8\: G_2$, e.g.\ $0.85 \approx 1.0-1/8$, and $1.4 \approx
1.5-1/8$. This is explained in Sec.~3 by the BCF model, but we 
summarize the conditions for plateau formation now. The width of a
transition is governed by a parameter $T^*$ being a combination of the
actual temperature $T$ and the tunneling parameter $T_t$. A normal
plateau occurs when the chemical potentials $\mu_s$ and $\mu_d$ are
more than a few times $k T^*$ away from subband edges $\ve_\alpha$. 
According to the BCF model an anomalous plateau occur when the
chemical potential $\mu_d$ of the drain is less than a few times $k
T^*$ away from an anomalous subband edge $\ve_\alpha'(\mu_d)$. In
short, we have the following conditions for plateau formation:

\begin{eqnarray}
\label{norm_plat}
{\rm normal\; plateau:}&
|\mu_{s,d} - \ve_\alpha| & >\; 4kT^*,\\
\label{anom_plat}
{\rm anomalous\; plateau:}&
|\mu_d - \ve'_\alpha(\mu_d)| & <\; 4kT^*.
 \end{eqnarray}
In Ref.~\cite{Bruus01} the function name $\Delta$ was introduced 
for the energy difference in Eq.~(\ref{anom_plat}), and when it is
positive it can be interpreted as the Fermi energy of the anomalous
subband, e.g.\ for $\ve'_0$
\begin{equation} \label{Delta_def}
\Delta(\mu_d) \equiv \mu_d - \ve'_n(\mu_d).
\end{equation}

We are now ready to determine the explicit expressions for the
subband edges based on the data in Fig.~\ref{fig:grayscaleEXP}. The
normal transition lines $\mu_s = \ve_1$ and $\mu_d = \ve_0$ cross at
the point $(V_{\rm gs},\: V_{\rm sd}) = (320~{\rm mV},\: 6.5~{\rm mV})$. 
Adding Eqs.~(\ref{transition_s}) and~(\ref{transition_d}) after insertion
of these numbers leads to $\ve_1-\ve_0 = eV_{\rm sd} = 6.5$~meV. If we
choose the energy zero to be at $\ve_0$ we end up with the following
normal subband edges.
\begin{eqnarray}
\label{e0}
\ve_0 &=& 0.0~{\rm meV},\\
\label{e1}
\ve_1 &=& 6.5~{\rm meV}.
\end{eqnarray}
Combining Eqs.~(\ref{VaVb}), (\ref{mu0_V}), 
(\ref{mu0amu0b}), (\ref{e0}) and (\ref{e1})  leads to
\begin{equation} \label{mu0_Vgs}
\mu_0(V_{\rm gs}) = 0.13\: V_{\rm gs} - 38.35~{\rm meV}.
\end{equation}
Concerning the anomalous $G=0.85\: G_2$ plateau in 
Fig.~\ref{fig:grayscaleEXP} we note that the white transition lines on
either side of it are almost parallel and separated by 
$\Delta V_{\rm gs} \approx 15$~mV. Using Eq.~(\ref{mu0_Vgs}) we can
deduce the width in energy space of the anomalous plateau to be
$\sim2$~meV. Combined with Eq.~(\ref{anom_plat}) we arrive at
\begin{equation} \label{ep0_approx}
\ve'_0(\mu_d)  \left\{
\begin{array}{ll}
\approx \mu_d,& {\rm for}\; \: 0<\mu_d<2~{\rm meV},\\
   =        0,& {\rm otherwise}.
\end{array} \right.  
\end{equation}

The three subband edges $\ve_0$, $\ve'_0(\mu_d)$, and $\ve_1$ are the
main input to the BCF model calculations used to analyze all our
experiments in the rest of the paper. 
\vspace*{3mm}

\noindent
{\bf 3. The BCF model at finite bias}\\[2mm]
The central idea in the BCF model is that the degeneracy of the first,
but in principle any, two-fold degenerate subband, is lifted as the
chemical potential of the drain crosses the subband edge $\ve_0$. One
of the resulting non-degenerate subband edges, the normal one, remains
at $\ve_0$ while the other, the anomalous $\ve'_0(\mu_d)$, follows
$\mu_d$ such that the deviation 
$\Delta(\mu_d) = \mu_d-\ve'_\alpha(\mu_d)$ is a power law
\cite{Bruus01}.  All the phenomenology of the 0.7 anomaly is contained
in this power law that leads to the all important tendency of the
anomalous subband edge to pin to $\mu_d$. We will not engage in a
discussion concerning the foundation of this model, but just use
it as a phenomenological input to the Landauer-B\"uttiker formalism,
and from this obtain a unified interpretation of our experimental
results. 

\begin{figure}[t]
\mbox{} \hfill  
\includegraphics[width=\columnwidth]{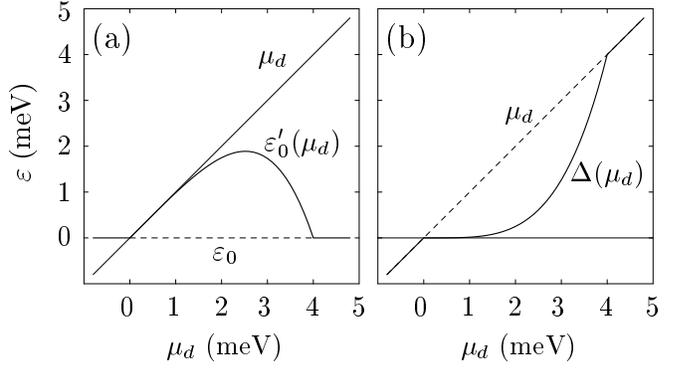}\\
\caption{ \label{fig:anom_edge_Delta} 
(a) The position of the anomalous subband edge $\ve'_0(\mu_d)$
given by Eq.~(\ref{ep0}) with $n=3$ and $\mu^* = 4.0$~meV as in 
Secs.~4 and 5. (b) Graph of the anomalous subband Fermi energy
$\Delta(\mu_d)$ Eq.~(\ref{Delta_def}) exhibiting the power
law behavior for $\mu_d$ near 0 mV required by the BCF model.
}  
\end{figure}

As summarized in Eq.~(\ref{ep0_approx}) the first anomalous conductance
plateau of sample A is determined experimentally to occur for
$0<\mu_d<2$~meV. In this interval we must demand that $\ve'_0(\mu_d)
\approx \mu_d$. If we also assume the power law behavior of the BCF
model, we can fulfill both these requirements by the following simple
function 
\begin{equation} \label{ep0}
\ve'_0(\mu_d) = \left\{
\begin{array}{cl}
\mu_d(1-\mu_d/\mu^*)^n,& {\rm for}\; \: 0<\mu_d<\mu^*,\\
0,& {\rm otherwise}.
\end{array} \right.  
\end{equation}
In Fig.~\ref{fig:anom_edge_Delta} are depicted the graphs of
$\ve'_0(\mu_d)$ and $\Delta(\mu_d)$ with the choice of parameters
$n=3$ and $\mu^* = 4.0$~meV as described in Sec.~4.

To be able to analyze Fig.~\ref{fig:grayscaleEXP} we extend the BCF
model, treated so far only in the linear response regime, by combining
it with the standard half-plateau theory \cite{Glazman89} based on the
finite bias version of the Landauer-B\"uttiker formula. Crucial
ingredients in this treatment are the subband edges $\ve_\alpha$ 
of the point contact, $\alpha = 0,0',1,\ldots$, and the energy
dependent transmission coefficients 
${\cal T}[\ve,\ve_\alpha]$. In practice these coefficients
cannot be determined completely, and here we restrict our treatment to
two simple examples, the perfect adiabatic transmission   
${\cal T}_{\rm ad}[\ve,\ve_\alpha]$ and the slightly more
realistic case \cite{Buttiker90} of non-adiabatic tunneling smeared
transmission ${\cal T}_{\rm tn}[\ve,\ve_\alpha]$:  
\begin{eqnarray} 
\label{Tad}
{\cal T}_{\rm ad}[\ve,\ve_\alpha(\mu_d)]
&=& \Theta[\ve-\ve_\alpha(\mu_d)],\\[2mm]
\label{Ttn}
{\cal T}_{\rm tn}[\ve,\ve_\alpha(\mu_d)]
&=& \frac{1}{\exp[(\ve_\alpha(\mu_d)-\ve)/k T_t]+1},
\end{eqnarray}
where $\Theta[x]$ is the usual step function, and $T_t$ is a parameter
(here written as a characteristic temperature) describing the degree
of smearing of the transmission coefficient due to reflection above
and tunneling below the subband edge. 

\begin{figure}[tbp]
\centerline{
  \includegraphics[width=\columnwidth]{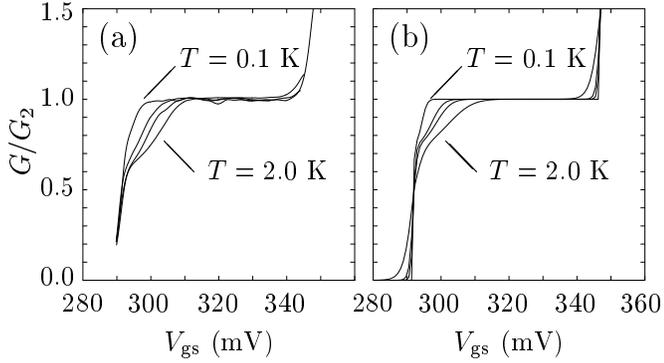}}
\caption{ \label{fig:GT} 
(a) The measured zero--bias differential conductance $G(V_{\rm gs})$
exhibiting a strengthening of the 0.7 anomaly as the temperatures is
raised from 0.1~K, through 0.5~K and 0.8~K, ending at 2.0~K.
(b) BCF model calculation  of $G_{\rm ad}$ Eq.~(\ref{G0_ad}) 
for $T =$ 0.1, 0.5, 0.8, and 2.0~K using the parameter values given in
Sec.~6. Note the fair agreement between experiment and theory even in
the simple case of adiabatic transmission. 
}  
\end{figure}

The current $I(V_{\rm sd},V_{\rm gs},T)$ at finite bias is calculated
in terms of the transmission coefficients, the degeneracy factor
$\lambda_\alpha=1,2$ of subband $\ve_\alpha$, and the Fermi-Dirac  
distribution function $f(\ve,\mu_{s/d},T)$ of the source and drain
reservoirs as
\begin{eqnarray}
\label{current_def} 
&& \hspace*{-7mm}
I\bigl[\mu_s(V_{\rm gs},V_{\rm sd}),
       \mu_d(V_{\rm gs},V_{\rm sd}),T\bigr]\\[2mm]
\nonumber 
&& \hspace*{-5mm}
= \frac{e}{h} \sum_\alpha \lambda_\alpha\!\!
\int_{-\infty}^{\infty}\!d\ve\:
{\cal T}\bigl[\ve,\ve_\alpha(\mu_d)\bigl]\:
\bigl[f(\ve,\mu_s,T)\!-\!f(\ve,\mu_d,T)\bigr].
\end{eqnarray}

Generally the current is found from Eq.~(\ref{current_def}) by
numerical integration. However, since $\int dx\: (e^x+1)^{-1} = x -
\ln(e^x+1)$, simple analytical results can be obtained when the
adiabatic transmission coefficient ${\cal T}_{\rm ad}$ of
Eq.~(\ref{Tad}) is used. In the following we shall use
${\cal T}_{\rm ad}$ extensively and only in Sec.~6 switch to the
non-adiabatic, tunneling smeared transmission coefficient 
${\cal T}_{\rm tn}$ of Eq.~(\ref{Ttn}). 

In the case of adiabatic transmission the finite bias, differential
conductance $G_{\rm ad}(V_{\rm gs},V_{\rm sd},T) = dI/dV_{\rm sd}$  
is 
\begin{eqnarray}
\label{G_ad}
&& \hspace*{-8mm}
G_{\rm ad} \bigl[\mu_s(V_{\rm gs},V_{\rm sd}),
       \mu_d(V_{\rm gs},V_{\rm sd}),T\bigr] =
\frac{G_2}{4}\: \sum_\alpha \lambda_\alpha\\
\nonumber
&& \hspace*{-5mm} \times
\left(
  \frac{\displaystyle 1}{\displaystyle 1 +  
  \exp \Bigl[ \frac{
  \displaystyle \ve_\alpha(\mu_d)\!-\!\mu_s}{
  \displaystyle kT} \Bigr] }
 + 
  \frac{\displaystyle 1}{\displaystyle 1 +  
  \exp \Bigl[ \frac{
  \displaystyle \ve_\alpha(\mu_d)\!-\!\mu_d}{
  \displaystyle kT} \Bigr] }
\right).
\end{eqnarray}
In the limit of zero temperature the conductance $G_{\rm ad}$
Eq.~(\ref{G_ad}) is seen to increase in units of 
$(\lambda_\alpha/4)\: G_2$ each time 
$\mu_s$ or $\mu_d$ crosses a subband edge $\ve_\alpha$. In the standard
case, where all subbands two--fold degenerated and all subband
edges are constant, the conductance unit is the usual $G_2$ if 
$V_{\rm sd} = 0$, while the half-plateau values $0.5\: G_2$ are
recovered for finite bias, $V_{\rm sd} \neq 0$. 

Anomalous plateaus can also be obtained from
Eq.~(\ref{G_ad}). Assuming the BCF model Eq.~(\ref{ep0})
the inequality $\ve'_0(\mu_d)\!<\!\mu_d  \ll kT$ is fulfilled in a
finite interval $0<\mu_d\ll\mu^*$, thus yielding
$1/(1+\exp[(\ve'_0(\mu_d)\!-\!\mu_d)/kT]) \approx 1/2$, i.e.\ half of
the zero temperature value. Since in the BCF model the degeneracy of
the lowest subband is lifted, we have $\lambda_{0'} = 1$. This means
that the conductance on the anomalous plateau is lowered by 
$1/4\: G_2$ if $V_{\rm sd} =  0$ (where $\mu_s =    \mu_d$) and by
$1/8\; G_2$ if $V_{\rm sd}\neq0$ (where $\mu_s \neq \mu_d$). 
Consequently, in accordance with the measurements, we find at zero bias
the anomalous plateau $G = (1-1/4)\: G_2 = 0.75\: G_2$, and at finite
bias $G = (1-1/8)\: G_2 = 0.875\: G_2$ as well as 
$G = (3/2-1/8)\: G_2 = 1.375\: G_2$.

As an example of a full conductance formula involving anomalous
plateaus we give the BCF formula for $G_{\rm ad}$ in the zero bias
limit where $\mu_s = \mu_d = \mu$. We restrict the formula to the
lowest three subbands $\ve_0$, $\ve'_0(\mu)$, and $\ve_1$ with
degeneracies $\lambda_0=1$, $\lambda_{0'}=1$, and $\lambda_1=2$,
respectively:  
\begin{eqnarray}
\label{G0_ad}
&& \hspace*{-7mm}
G_{\rm ad}(\mu,\mu,T) = G_2 \\
&& \hspace*{-5mm} \nonumber
\times \Biggl(
       \frac{1/2}{1\!+\!e^{(\ve_0      -\mu)/k T}}
 \!+\! \frac{1/2}{1\!+\!e^{(\ve'_0(\mu)-\mu)/k T}}
 \!+\! \frac{  1}{1\!+\!e^{(\ve_1      -\mu)/k T}} \Biggr),
\nonumber
\end{eqnarray}
In Fig.~\ref{fig:GT} is shown a comparison at different temperatures
between low-bias experimental measurements on $G$ and BCF model
calculations based on Eq.~(\ref{G0_ad}). A good qualitative agreement
is seen, in particular concerning the strengthening of the 0.7
conductance anomaly as the temperature is increased. In Sec.~6 we
give a more detailed analysis of the temperature dependence of the
0.7 conductance anomaly at zero bias.
\vspace*{3mm}

\begin{figure}[ht]
\centerline{\includegraphics[height=80mm]{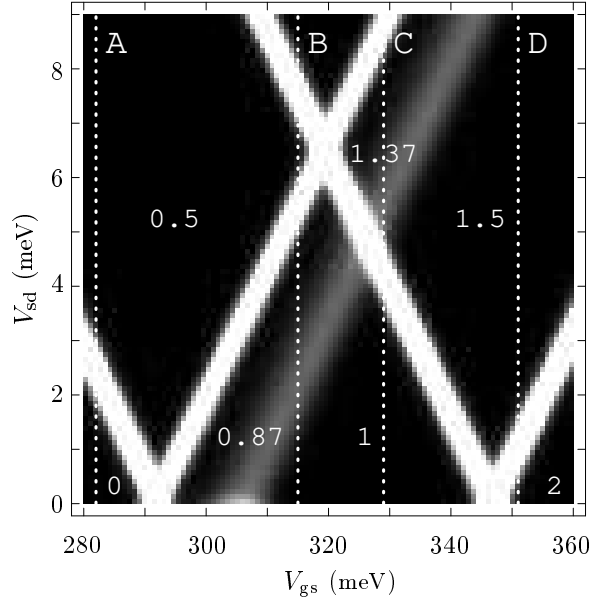}}
\caption{\label{fig:grayscaleBCF}
Calculated transconductance $dG/dV_{\rm gs}$ based on the
conductance formula Eq.~(\ref{G_ad}), valid in the case of adiabatic
transmission, using the BCF model of the anomalous subband edge
$\ve'_0(\mu_d)$ Eq.~(\ref{ep0}) with $n=3$ and $\mu^* = 4$~meV. To
mimic tunneling--induced smearing of the plateaus the temperature is
set to $1$~K, somewhat higher than the 0.3~K of the actual
experiment. The sweeps in $V_{\rm sd}$ along the lines A--D are
studied in Sec.~5. 
}
\end{figure}

\noindent
{\bf 4. Transconductance in the BCF model}\\[2mm]
By differentiation of Eq.~(\ref{G_ad}) it is a simple matter to
calculate the transconductance $dG/dV_{\rm gs}$. This can then be
used to reproduce the grayscale plot of Fig.~\ref{fig:grayscaleEXP}
theoretically by combining it with the experimentally determined
subband edges $\ve_0$, Eqs.~(\ref{e0}), and $\ve_1$, Eq.~(\ref{e1}) as
well as the BCF model of the anomalous edge $\ve'_0(\mu_d)$, 
Eq.~(\ref{ep0}). For the latter we use the values $n=3$ and 
$\mu^* = 4.0$~meV found by visual fitting to the data. The resulting
grayscale plot of the transconductance is shown in
Fig.~\ref{fig:grayscaleBCF}. 

The main features are well reproduced: the straight white lines of the
half-plateau model and the gray line with positive slope parallel to
the white line forming the right-hand edge of the $G=0.5\: G_2$
plateau. A number of minor features are not reproduced at all, simply
due to the omission of certain physical effects in the model. 
\\[-5mm]

\begin{figure}[ht]
\centerline{
     \includegraphics[width=\columnwidth]{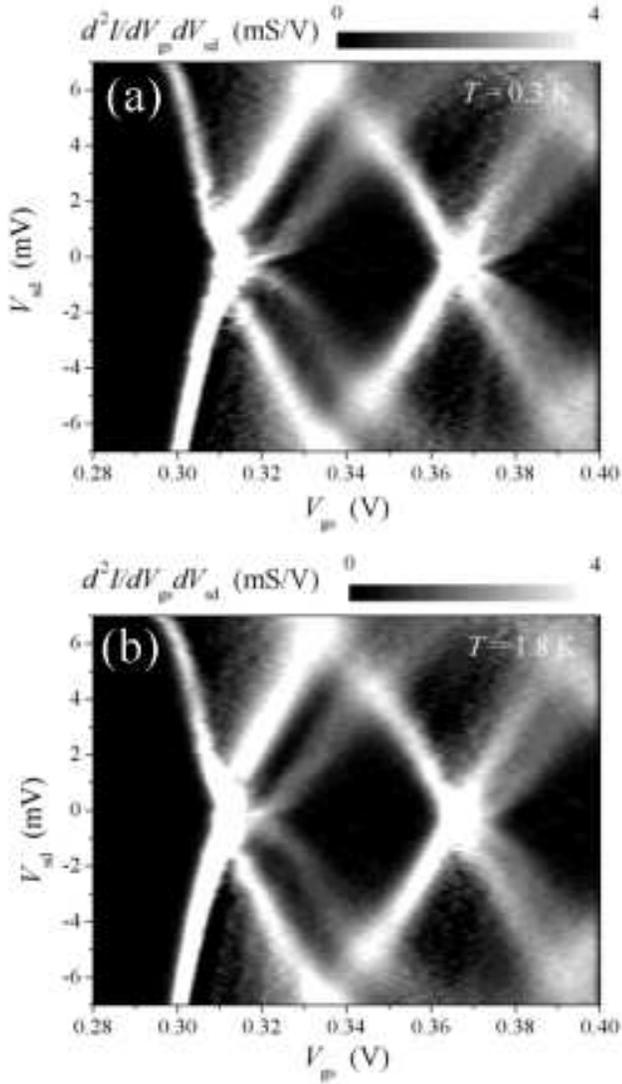}}
\caption{\label{fig:grayscale_temp}
Differential transconductance as in Fig.~\ref{fig:grayscaleEXP} 
measured at $T = 0.3$ and 1.8~K. The structures are not changed
significantly by changing the temperature by a factor of six.
}
\end{figure}

Firstly, the left border of the experimental
$G=0.5\: G_2$ plateau is not quite straight. This is presumably due to
Coulomb interaction effects dominating the point contact at the very
low densities as it opens up for conduction. Electron interaction was
not included in the model.  

Secondly, most structures in Fig.~\ref{fig:grayscaleEXP} tend to be
smeared out at high values of $V_{\rm sd}$. This is probably due to
heating of the electron system by applying such high values of the
bias voltage. Also, the Fermi energy of the 2DEG reservoirs is 10~meV,
and when $eV_{\rm sd}$ at the top of the grayscale plot approaches
this value, strong distortions of the system occur. Neither of these
two effects are taken into account in the model calculation.

Thirdly, the anomalous 0.85--1 transition line is not quite parallel
to the white 0.5--0.85 transition line in the experiment. This may
also be an effect of bias induced heating. As the temperature is
increased the width of the anomalous plateau is increased as shown by
the zero--bias data in Fig.~\ref{fig:GT}. This would lead to an
increased distance between the 0.5--0.85 and 0.85--1.0 transition
line as $V_{\rm sd}$ is increased. That heating due to the finite bias
is important may be deduced from  Fig.~\ref{fig:grayscale_temp}. In
this figure are shown two experimental grayscale plots recorded at the
two cryostat temperatures 0.3~K and 1.8~K, respectively. The main
features of the plots are not changed significantly by the six--fold 
increase of the temperature. However, except for the 0--0.5
transition line, there is the above mentioned tendency of stronger
smearing of the transition lines as $V_{\rm sd}$ is increased. Thus
heating due to the cryostat temperature is negligible compared to the
smearing due to bias induced heating.

Lastly, at low bias the anomalous 0.85--1 transition line deviates
from a straight line in the experiment: the width of the anomalous
plateau shrinks as the bias is lowered. In the model calculation we
have used constant values for the parameters $\mu^*$ and $n$ in
Eq.~(\ref{ep0}), and therefore we obtained a constant width.

Further studies of all these effects are in progress.
\vspace*{3mm}

\begin{figure}[ht]
\hfill \includegraphics[width=\columnwidth]{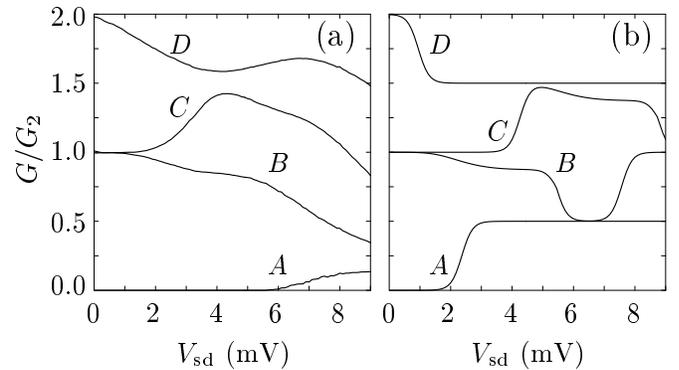}\\
\caption{\label{fig:Ggraf}
(a) The measured differential conductance $G(V_{\rm sd})$ along the
four $V_{\rm sd}$ sweep-lines A--D indicated in
Fig.~\ref{fig:grayscaleEXP} with $T = 0.3$~K. (b) BCF model
calculation of $G(V_{\rm sd})$ along the four $V_{\rm sd}$ sweep-lines
A--D indicated in Fig.~\ref{fig:grayscaleBCF} using Eq.~(\ref{G_ad})
with adiabatic transmission ${\cal T}_{\rm ad}$ and $T = 1.0$~K.
}
\end{figure}

\noindent
{\bf 5. \boldmath{$G$} at finite bias, experiments and model}\\[2mm]
The grayscale plots of the transconductance in
Figs.~\ref{fig:grayscaleEXP} and~\ref{fig:grayscaleBCF} expose clearly
the position of the subband edges but reveals nothing about the value
of the conductance at the plateaus. To find the plateau values we go
back and study the differential conductance $G$. In
Fig.~\ref{fig:Ggraf} is shown the graphs of $G(V_{\rm sd})$ obtained
from the bias sweeps A--D indicated in the experimental and
theoretical grayscale plots Figs.~\ref{fig:grayscaleEXP}
and~\ref{fig:grayscaleBCF}. 

The experimental curves of $G(V_{\rm sd})$ in Fig.~\ref{fig:Ggraf}(a) 
recorded for $V_{\rm gs} =$ 284, 317, 334, and 358~mV shows the
appearance of the half-plateaus 0.5, 1.0, 1.5 and 2.0, but they are
in particular chosen to show the quite well developed anomalous
plateau $G = 0.85\: G_2$, and the rather smeared anomalous plateau
at $G = 1.35\: G_2$.

The BCF model calculation of $G(V_{\rm sd})$ in Fig.~\ref{fig:Ggraf}(b) 
with $V_{\rm gs} =$ 282, 315, 329, and 351~mV reproduces the
experimentally observed features fairly well. In particular well
developed anomalous plateaus are seen at $G=$ 0.87 and $1.37\: G_2$. 

\vfill

\pagebreak

Naturally, the remarks made in Sec.~4 concerning the deficiencies of
the BCF model are also valid here. The experimentally observed bending
of the 0--0.5 transition line in the grayscale plot shows up as a
large upward shift in $V_{\rm sd}$ of the D--curve in
Fig.~\ref{fig:Ggraf}(a) as compared to the D--curve in the model
calculation Fig.~\ref{fig:Ggraf}(b). The strong deformation of the
spectrum for high bias voltages in the experiment appears as a marked
downward bending of the A, B, and C--curve in
Fig.~\ref{fig:Ggraf}(a). Finally, we also see the effect of the absence of
tunneling in the model calculation: the calculated transitions between
plateaus in Fig.~\ref{fig:Ggraf}(b) are much steeper than the observed
ones in Fig.~\ref{fig:Ggraf}(a) despite the fact that the temperature in
the model is set to 1.0~K, i.e.\ higher than the 0.3~K of the experiment.
\vspace*{3mm}

\noindent
{\bf 6. Activated conductance suppression at low bias}\\[2mm]
The fair agreement between experiment and model calculation in
Fig.~\ref{fig:GT} makes a comparison desirable which is more detailed
than the simplistic Arrhenius analysis presented in
Ref.~\cite{Kristensen00}. As mentioned in the end of Sec.~4 the width
of the plateau shrinks as $V_{\rm sd}$ goes to zero. We therefore use
the zero bias data in Ref.~\cite{Kristensen00} to fit the values for 
the parameters $n$ and $\mu^*$ in Eq.~(\ref{ep0}), and we find
$n=1$ and $\mu^* = 8.5$~meV. Due to the exponential factors appearing
in the Fermi-Dirac distribution functions it is natural to plot the
deviation from perfect quantization as $\ln[1 - G(T)/G_2]$ versus
$1/T$. We set $V_{\rm sd} = 0$~mV and focus on the first half of the
first plateau by choosing the following four fixed values of the
gate--source voltage: $V_{\rm gs} =$ 305, 310, 315, or 320~mV.

\begin{figure}[ht]
\centerline{
  \includegraphics[width=\columnwidth]{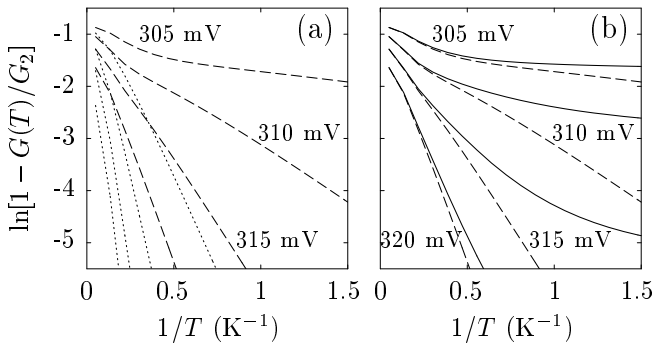}}
\caption{\label{fig:actNormalBCF} 
(a) Comparison of the calculated conductance suppression
$\ln[1-G(T)/G_2]$ versus $1/T$ with adiabatic transmission 
for $V_{\rm gs} =$ 305, 310, 315, and 320~mV in two cases. Dotted
lines: normal rigid subband edges, and dashed lines: the anomalous
density dependent subband edge, i.e.\ the BCF-model. (b) The BCF-model
as in (a) with adiabatic transmission (dashed lines) compared to the
BCF-model with the non-adiabatic tunneling transmission of
Eq.~(\ref{Ttn}) (full lines).
} 
\end{figure}

In Fig.~\ref{fig:actNormalBCF}(a) we compare model calculations based
on the conduction Eq.~(\ref{G_ad}) for the case of adiabatic
transmission ${\cal T}_{\rm ad}$ of Eq.~(\ref{Tad}) in the BCF
model and for the case of normal rigid subband edges fixed at $\ve_0$
and $\ve_1$. We see how the two calculations deviates both for small
and large values of $1/T$. The BCF values are systematically higher
than the values of the rigid subband edge model.

\vfill

\pagebreak

\noindent
In Fig.~\ref{fig:actNormalBCF}(b) we compare two versions of the BCF
model: one, the adiabatic case from Fig.~\ref{fig:actNormalBCF}(a)
using ${\cal T}_{\rm ad}$ of Eq.~(\ref{Tad}), and the other using the
non-adiabatic tunneling transmission  ${\cal T}_{\rm tn}$ of
Eq.~(\ref{Ttn}) with $T_t=1.1$~K. Here we note that in the high
temperature regime (small values of $1/T$) the two cases produce
identical results: tunneling is not important. However, at
temperatures lower than 1~K (large values of $1/T$), substantial
deviations occur due to tunneling induced suppression of the
conductance: the non-adiabatic point contact is further away from
perfect quantization, hence resulting in higher values in the plot.

\begin{figure}[tbp]
\centerline{\includegraphics[]{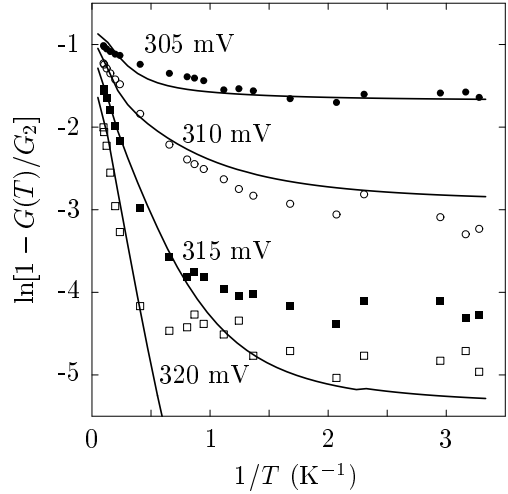}}
\caption{\label{fig:Arrhenius}
Comparison of experimental data (points) and BCF model calculation 
(full lines) of the conductance suppression $\ln[1-G(T)/G_2]$ versus
$1/T$ for the gate-­source voltages $V_{\rm gs} =$ 
305 ({\Large $\bullet$}), 310 ({\Large $\circ$}), 
315 ($\blacksquare$), and 320~mV ($\square$).
The data are obtained from the same sample as in
Figs.~\ref{fig:gatebias_surface} and \ref{fig:grayscaleEXP}. The
calculation is performed using the non-adiabatic tunneling
transmission ${\cal T}_{\rm tn}$ Eq.~(\ref{Ttn}) with the smearing
parameter $T_t = 1.1$~K. 
}
\end{figure}

Finally, in Fig.~\ref{fig:Arrhenius} we compare experimental values of
$\ln[1-G(T)/G_2]$ with the BCF model calulation of
Fig.~\ref{fig:actNormalBCF}(b) including smearing by 
tunneling. Good agreement is obtained in the high temperature
regime in all four cases, while in the low temperature regime (large
values of $1/T$) only the most strongly deviating conductance traces
are comparing well. It is to be expected that when studying minute
deviations from perfect quantization the exact form of 
${\cal T}[\ve,\ve_\alpha]$ plays an increasingly important role. There
is a priori no reason why  ${\cal T}_{\rm tn}[\ve,\ve_\alpha]$ of
Eq.~(\ref{Ttn}) should match the actual transmission function of the
sample. 
\vspace*{3mm}

\noindent
{\bf 7. Conclusion}\\[2mm]
In this work we have shown how the experimental determination at
finite bias and low temperatures of the three subband edges $\ve_0$,
$\ve'_0(\mu_d)$, and $\ve_1$ can be used as input to the BCF model
resulting in model calculations, which can predict the results of other
experiments. In particular we have shown this to be the case
explicitly for grayscale plot of the transconductance, the finite bias
values of the differential conduction $G(V_{\rm sd})$, and of the
temperature dependence of the suppression $\ln[1-G(T)/G_2]$. In all
cases the model calculations resulted in fair agreement with the
experiments, and we have thereby demonstrated how the BCF model with
the subband edges as input provides a unified way to analyze and
understand a wide range of data related to the 0.7 anomaly.

This way of analyzing experiments related to the 0.7 anomaly may be
very profitable to apply to the existing data on other experiments
than those treated in this paper: the series conductance of two 
point contacts \cite{Liang99}, suppression of shot-noise
\cite{Griffiths00}, thermopower measurements \cite{Appleyard} and
acousto-electric currents induced by surface acoustic waves (SAW)
\cite{Cunningham2001}. For the thermopower, a BCF model calculation
comparing well with experiments has already been performed successfully
\cite{Bruus01b}. 

This and previous work has led us to believe that the anomalous
subband edges combined with the BCF model may be the key to a deeper
understanding of the 0.7 structure. To allow for better analysis along
these lines we would therefore like to encourage experimentalists
always to measure the subband edges, using e.g.\ the finite bias
transconductance, whenever measuring other effects related to
the 0.7 anomaly. Finally, we note that at present 
the BCF model has no microscopic foundation, and naturally it is highly
desirable to derive such a foundation theoretically.
\vspace*{3mm}

\noindent
{\bf Acknowledgements}\\[2mm]
This work was supported by the Danish Technical Research Council
(grant no.\ 9701490) and by the Danish Natural Science Research Council
(Ole R{\o}mer grant no.\ 9600548). The GaAs heterostructure used in
this investigation was grown at the III-V NANOLAB, Ørsted Laboratory,
Niels Bohr Institute, by Claus B.\ S{\o}rensen.
\vspace*{3mm}

\noindent
{\bf References}\\[-15mm]

\normalsize
{\small
\setlength{\baselineskip}{4mm}

}


\begin{thebibliography}{99}

\bibitem{vanWees88} B.J. van Wees, H. van Houten, C.W.J.  Beenakker,
  J.G. Williamson, L.P. Kouwenhoven, D. van der Marel, and C.T. Foxon,
  Phys.Rev.Lett. {\bf B 60}, 848 (1988).

\bibitem{Wharam88} D.A. Wharam, T.J. Thornton,R. Newbury, M. Pepper,
  H. Ahmed, J.E.F. Frost, D.G. Hasko, D.C. Peacock, D.A. Ritchie,
  and G.A. Jones, J.Phys.C {\bf 21}, L209 (1988).

\bibitem{vanHouten92} H. Van Houten, C.W.J.Beenakker, and B. van Wees,
  p. 9 in {\em Nanostructured Systems, M. Reed eds.},
  {\em Semiconductors and Semimetals, R.K. Williamson, A.C. Beer and
  R. Weber eds.} (Academic Press, 1992).

\bibitem{Thomas96} K.J. Thomas, J.T. Nicholls, M.Y. Simmons, M. Pepper,
  D.R. Mace, and D.A. Ritchie, Phys. Rev. Lett. {\bf 77}, 135 (1996).

\bibitem{Tscheuschner96} R.D. Tscheuschner and A.D. Wieck,
  Superlattices and Microstructures {\bf 20}, 615 (1996).

\bibitem{Kristensen98a} A. Kristensen, J.B. Jensen, M. Zaffalon,
  C.B. S{\o}rensen, S.M. Reimann, P.E. Lindelof, M. Michel, and
  A. Forchel,
  J. Appl. Phys. {\bf 83}, 607 (1998).

\bibitem{Kristensen98b} A. Kristensen, P.E.Lindelof, J.B. Jensen,
  M. Zaffalon, J. Hollinghery, S.W. Pedersen, J.  Nyg{\aa}rd, H. Bruus,
  S.M. Reimann, C.B. S{\o}rensen, M. Michel, and A. Forchel,
  Physica B {\bf 249-251}, 180 (1998).

\bibitem{Kane98} B.E. Kane, G.R. Facer, A.S. Dzurak, N.E. Lumpkin,
  R.G. Clark, L.N. Pfeiffer, and K.W. West,
  Appl. Phys. Lett. {\bf 72}, 3506 (1998).
         
\bibitem{Thomas00} K.J. Thomas, J.T. Nicholls, M. Pepper, W.R. Tribe,
  M.Y. Simmons, and D.A. Ritchie, Phys. Rev. B {\bf 61}, R13365 (2000).

\bibitem{Kristensen00} A. Kristensen, H. Bruus, A. Forchel,
  J.B. Jensen, P.E. Lindelof, M. Michel, J. Nyg{\aa}rd, and
  C.B. S{\o}rensen, Phys. Rev. B {\bf 62}, 10950 (2000).

\bibitem{Pyshkin00} K.S. Pyshkin, C.J.B. Ford, R.H. Harell, M.
Pepper, E.H. Linfield, and D.A. Ritchie, Phys. Rev. B {\bf 62},
15842 (2000).


\bibitem{Nuttinck} S. Nuttinck, K. Hashimoto, S. Miyashita, 
T. Saku, Y. Yamamoto, and Y. Hirayama. 
Jpn. J. Appl. Phys. {\bf 39}, L655 (2000).

\bibitem{Hashimoto} K. Hashimoto, S. Miyashita, T. Saku, and Y. Hirayama
Jpn. J. Appl. Phys. {\bf 40}, 3000 (2001).

\bibitem{Reilly00} D. J. Reilly, G. R. Facer, A. S. Dzurak,
  B. E. Kane, R. G. Clark, P. J. Stiles, J. L. O'Brien, N. E. Lumpkin,
  L. N. Pfeiffer, K. W. West, 
  Phys. Rev. B {\bf 63}, R121311 (2001).

\bibitem{Thomas98b} K.J. Thomas, J.T. Nicholls,M.Y. Simmons,
  M. Pepper, D.R. Mace, and D.A. Ritchie, 
  Phil. Mag. {\bf 77}, 1213 (1998)

\bibitem{Thomas98} K.J. Thomas, J.T. Nicholls, N.J. Appleyard,
  M. Pepper, M.Y. Simmons, D.R. Mace, W.R. Tribe and D.A. Ritchie,
  Phys. Rev. B {\bf 58}, 4846 (1998).

\bibitem{Liang99} C.-T. Liang, M.Y. Simmons, C.G. Smith, G.H. Kim,
  D.A. Ritchie and M. Pepper,
  Phys. Rev. B {\bf 60}, 10687 (1999).

\bibitem{Griffiths00}
  T.G.\ Griffiths , E. Comforti, M. Heiblum, and V. Umansky, ICPS-25
  workbook, 316 (2000). 

\bibitem{Appleyard}
  N. J. Appleyard, J. T. Nicholls, M. Pepper, W. R. Tribe,
  M. Y. Simmons, and D. A. Ritchie,
  Phys. Rev. B {\bf 62}, R16275 (2000).

\bibitem{Cunningham2001} J. Cunningham, V.I. Talyanskii, J.M.
Shilton, M. Pepper, A. Kristensen, and P.E. Lindelof, to be
published (2001).

\bibitem{Bruus01} H. Bruus, V.V. Cheianov, and K. Flensberg,
  Physica E {\bf 10}, 97-102 (2001).


\bibitem{Patel91} N.K. Patel, J.T. Nicholls, L. Martin-Moreno,
  M. Pepper, J.E.F. Frost, D.A. Ritchie and G.A.C. Jones
  Phys. Rev. B {\bf 44}, 13549 (1991).

\bibitem{Glazman89} L.I. Glazman and A. V. Khaetskii,
  Europhys. Lett. {\bf 9}, 263 (1989).

\bibitem{Buttiker90} M. B\"uttiker, 
  Phys. Rev B {\bf 41}, 7906 (1990) 

\bibitem{Bruus01b} H. Bruus, V.V. Cheianov, and K. Flensberg,
Proc. XXXVI Rencontres de Moriond (in press, 2001).

\end{thebibliography}
\end{document}